\documentclass[letterpaper, 10 pt, conference]{IEEEtran}
\IEEEoverridecommandlockouts
\usepackage{cite}
\usepackage{amsmath,amssymb,amsfonts}

\DeclareMathOperator*{\argmin}{arg\,min}
\usepackage{algorithmic}
\usepackage{graphicx}
\usepackage{textcomp}
\usepackage{xcolor}
\usepackage{bm}
\usepackage[english]{babel}
\graphicspath{{figures/}}

\def\BibTeX{{\rm B\kern-.05em{\sc i\kern-.025em b}\kern-.08em
    T\kern-.1667em\lower.7ex\hbox{E}\kern-.125emX}}
\usepackage[hmargin=0.75in, vmargin=0.75in]{geometry}
\usepackage{afterpage}
\usepackage{booktabs}
\usepackage{multirow}
\usepackage{threeparttable}

\begin{document}
\newgeometry{hmargin=0.75in, top=1in, bottom=0.75in}

\title{Scaling up the Optimal Safe Control of Connected and Automated Vehicles to a Traffic Network: A Hierarchical Framework of Modular Control Zones\\
    \thanks{%
    This work was supported in part by NSF under grants ECCS-1931600,
    DMS-1664644 and CNS-1645681, by ARPAE under grant DE-AR0001282, by AFOSR
    under grant FA9550-19-1-0158, and by the MathWorks. }
    \thanks{K. Xu and C. G. Cassandras are with the
    Division of Systems Engineering and Center for Information and Systems
    Engineering, Boston University, Brookline, MA, 02446, USA \{xky, cgc\}@bu.edu}
    }
	\author{\IEEEauthorblockN{Kaiyuan Xu} \and \IEEEauthorblockN{Christos G. Cassandras} }
\maketitle
\afterpage{\afterpage{\aftergroup\restoregeometry}}

\begin{abstract}
We consider the problem of scaling up optimal and safe controllers for Connected and Automated Vehicles (CAVs) from a single Control Zone (CZ) around a traffic conflict area to an entire network. The goal is to jointly minimize travel time and energy consumption for all CAVs, while providing speed-dependent safety guarantees within a CZ and satisfying velocity and acceleration constraints. A hierarchical modular CZ framework is developed consisting of a lower level where decentralized controllers are used that
combine Optimal control and Control Barrier Functions (OCBF) and a higher level where a feedback flow controller is proposed to coordinate adjacent CZs. The flow controller is parameterized by a terminal velocity constraint that serves as the interface between CZs. Simulation results show that the proposed modular control zone framework outperforms a direct extension of the OCBF framework to multiple CZs without any flow control.

\end{abstract}


\section{Introduction}
The performance of transportation networks critically depends on the management of traffic at conflict areas such as intersections, roundabouts and merging roadways \cite{rios2016survey}. The objectives of coordinating and controlling vehicles in such conflict areas include 
reducing congestion and energy consumption while also ensuring passenger comfort and guaranteeing safety \cite{chen2015cooperative, tideman2007review}. The emergence of Connected
and Automated Vehicles (CAVs) \cite{rios2016survey} and the development of new traffic infrastructure technologies \cite{li2013survey} provide promising new solutions to this problem through better information utilization and more precise vehicle trajectory design.

Most research to date has focused on the control and coordination of CAVs within a single \emph{Control Zone} (CZ) that encompasses a conflict area and the space around it within which a single coordinator can maintain data and communication is feasible between CAVs (V2V) or between CAVs and the coordinator (V2I). Both centralized and decentralized methods have been used to deal with this problem. Centralized mechanisms are often invoked for forming platoons in merging problems \cite{xu2019grouping} and determining passing sequences at intersections \cite{xu2020bi}. These approaches tend to work better when the safety constraints are independent of speed and they generally require significant computation and communication resources, especially when traffic is heavy. They are also not easily amenable to disturbances. Decentralized mechanisms restrict all computation to be done on board each
CAV with information sharing limited to a small number of neighboring
vehicles \cite{milanes2010automated, rios2015online, bichiou2018developing,
hult2016coordination}. Optimal control problem formulations are often used, with Model Predictive Control (MPC) techniques employed as an alternative to account for additional constraints and to compensate for disturbances by re-evaluating optimal actions over receding horizons \cite{cao2015cooperative, mukai2017model, nor2018merging}. The objectives in such problem formulations typically target the minimization of acceleration or the maximization of passenger comfort (measured as the acceleration derivative or jerk). Alternatives to MPC have recently been proposed through the use of Control
Barrier Functions (CBFs) \cite{xiao2020bridging} which provide provable guarantees that safety constraints are always satisfied.

However, the transition from a single CZ to \emph{multiple} interconnected CZs
is particularly challenging for several reasons including the following. First, when studying a CZ in isolation through an optimization problem, it is assumed that the initial conditions for each CAV satisfy the constraints of the problem. When one CZ is followed by another, there is no obvious way to ensure that the state of a CAV exiting the first can always satisfy the initial feasibility conditions of the next CZ's optimization problem. 
Second, when two adjacent CZs are physically separated by a short distance, the optimal control of CAVs in either CZ may in fact cause congestion in the other, including traffic blocking effects.
Thus, both performance degradation and lack of safety guarantees result from a direct application of the techniques developed to date for a single CZ without utilizing the global information of the traffic flow in a system or at least some local information from neighboring CZs.

The contribution of this paper is to take a step towards a systematic extension of traffic control involving CAVs from one to multiple CZs and ultimately a complex traffic network. We propose a hierarchical control framework consisting of modular interconnected CZs with two levels: $(i)$ a decentralized lower level applied to each CZ, and $(ii)$ an upper traffic flow control level. The interaction between the two levels relies on
an interface that connects neighboring CZs and adjusts the values of velocity parameters controlled by the upper level to influence the performance of CAVs at the lower level, where CAVs are still controlled in a decentralized way. These velocity parameters effectively act as regulators of the traffic flow between adjacent CZs and can be dependent on global or only local information regarding the traffic network.
The decentralized controllers used at the lower level are based on our prior work that combines Optimal Control with Control Barrier Functions (OCBF) \cite{xiao2020bridging}.
In the OCBF framework, a decentralized optimal control problem is first formulated with an objective which jointly minimizes $(i)$ the travel time of each CAV over a given road segment from a point entering a Control Zone (CZ) to the eventual exit point, and $(ii)$ a measure of its energy consumption. CBF-based constraints are included to \emph{guarantee} that all safety constraints are satisfied at all times, taking advantage of the forward invariance property of CBFs \cite{xiao2023safe}.



The paper is organized as follows. In Section II, the structure of the hierarchical control framework is introduced and a merging control problem for CAVs in two consecutive control zones is formulated to illustrate its functionality. In Section III, explicit solutions are derived for both the lower level, consisting of OCBF controllers for CAVs in each CZ, and the upper flow control level. Simulation results in Section IV show the performance improvements obtained using the hierarchical framework for modular CZs compared to a direct extension from one to multiple CZs with no flow control coordination between CZs.

\section{Problem Formulation}
\begin{figure*}
    \centering
    \includegraphics[width=\linewidth]{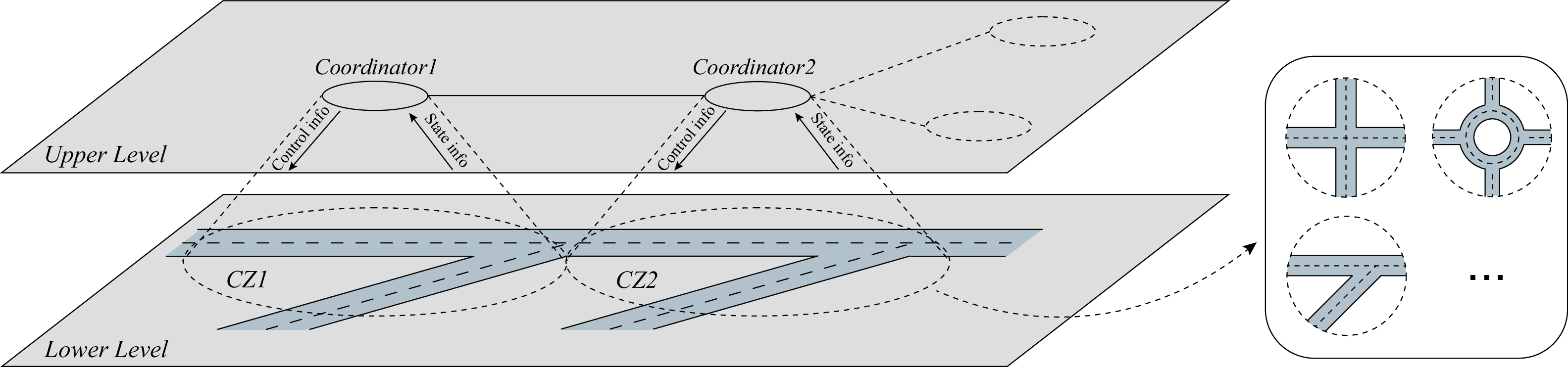}
    \caption{Modular Control Zone Architecture}
    \label{fig:module_CZ}
\end{figure*}

We model the control of CAVs in a general traffic network under the modular control zone architecture shown in Fig. \ref{fig:module_CZ}. A \emph{Control Zone} (CZ), defined as an area within which CAVs can communicate with each other and with a coordinator, 
is usually associated with a specific conflict area, including (but not limited to) merging roadways, intersections, and roundabouts. A coordinator (Road Side Unit (RSU)) is associated with each CZ to maintain a certain passing sequence of CAVs based on their arrival time at the CZ and their states; it also enables real-time Vehicle-to-Infrastructure (V2I) communication with the CAVs that are in the CZ. In prior work involving a single CZ \cite{xiao2020bridging}, the coordinator does not have any control functionality over individual CAVs and is only responsible for maintaining and sharing state information with the CAVs.
In the proposed modular CZ framework, a coordinator has the additional function of controlling traffic flows between CZs by sending information to CAVs based on the real-time CZ information acquired through Infrastructure-to-Infrastructure (I2I) communication. To maintain the \emph{decentralized} framework within each CZ, the signals sent by the coordinator to a CAV do not directly affect its control; 
they are instead parameters that influence the dynamics of each CAV (e.g., its target terminal velocity upon exiting the CZ). These signals also act as an interface between CZs to control their traffic flows.

In this paper, we limit ourselves to the merging control problem in consecutive CZs as shown in Fig. \ref{fig:module_CZ} so as to illustrate the framework of modular CZs. The problem is decomposed into $(i)$ a lower level with some modifications to the established decentralized merging control problem \cite{xiao2020bridging}, and $(ii)$ a new upper level with the flow control problem. These two levels are detailed next. 


\subsection{Merging Control: Lower Level Problem}
\begin{figure}[ptbh]
\centering
\includegraphics[width=\linewidth]{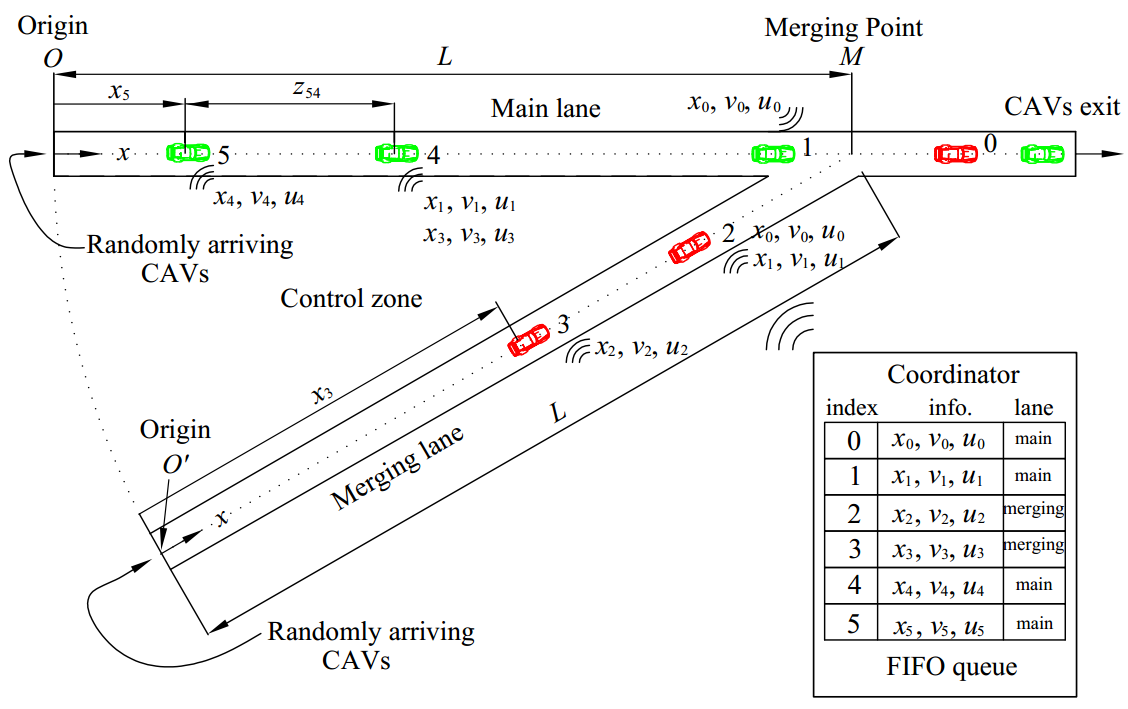} \caption{The merging problem:
{CAVs randomly arrive at the origins of the main and merging roads.
Collisons may occur at the Merging Point $M$. A coordinator is associated with $M$ to
maintain a queue and share information among CAVs as needed.}}%
\label{modelF}%
\end{figure}

We consider the merging problem arising when traffic must be joined from two different roads, usually associated with a main road and a merging road as shown in
Fig. \ref{modelF}. We consider the case where all traffic consists of CAVs
randomly arriving at the two roads joined at the \emph{Merging Point} (MP) $M$ where a collision may occur. A coordinator is associated with the MP to maintain a First-In-First-Out (FIFO) queue of CAVs based on their arrival time at the CZ. The segment from the origin $O$ or $O^{\prime}$ to $M$ has a length $L$ for both roads, where $L$ is selected to be as large as possible subject to the coordinator's communication range and the road network's
configuration. Since we consider single-lane roads in this merging problem, CAVs may not overtake each other in the CZ (extensions to multi-lane merging are given in \cite{XiaoITSC2020}).  The FIFO assumption imposed so that CAVs cross the MP in their order of arrival is made for simplicity (and often to ensure fairness), but can be relaxed through dynamic resequencing schemes as in \cite{XiaoACC2020} where optimal solutions are similarly derived but for different selected CAV sequences.

Let $S(t)$ be the set of FIFO-ordered indices of all CAVs located in the CZ at
time $t$ along with the CAV (whose index is 0 as shown in Fig.\ref{modelF})
that has just left the CZ. Let $N(t)$ be the cardinality of $S(t)$. Thus, if a
CAV arrives at time $t$, it is assigned the index $N(t)$. All CAV indices in
$S(t)$ decrease by one when a CAV passes over the MP and the vehicle whose
index is $-1$ is dropped.

The vehicle dynamics for each CAV $i\in S(t)$ along the lane to which it
belongs take the form
\begin{equation}
\left[
\begin{array}
[c]{c}%
\dot{x}_{i}(t)\\
\dot{v}_{i}(t)
\end{array}
\right]  =\left[
\begin{array}
[c]{c}%
v_{i}(t)\\
u_{i}(t)
\end{array}
\right]  \label{VehicleDynamics}%
\end{equation}
where $x_{i}(t)$ denotes the distance to the origin $O$ ($O^{\prime}$) along
the main (merging) lane if the vehicle $i$ is located in the main (merging)
lane, $v_{i}(t)$ denotes the velocity, and $u_{i}(t)$ denotes the control
input (acceleration). We consider two objectives for each CAV subject to four
constraints, as detailed next.

$\textbf{Objective 1}$ (Minimizing travel time): Let $t_{i}^{0}$ and
$t_{i}^{m}$ denote the time that CAV $i\in S(t)$ arrives at the origin $O$ or
$O^{\prime}$ and the MP $M$, respectively. We wish to minimize the travel time
$t_{i}^{m}-t_{i}^{0}$ for CAV $i$.

$\textbf{Objective 2}$ (Minimizing energy consumption): We also wish to
minimize energy consumption for each CAV $i\in S(t)$ expressed as
\begin{equation}
\setlength{\abovedisplayskip}{2pt}\setlength{\belowdisplayskip}{2pt}
J_{i}(u_{i}(t))=\int_{t_{i}^{0}}^{t_{i}^{m}}C(u_{i}(t))dt,
\end{equation}
where $C(\cdot)$ is a strictly increasing function of its argument.

$\textbf{Constraint 1}$ (Rear end safety constraints): Let
$i_{p}$ denote the index of the CAV which physically immediately precedes $i$
in the CZ (if one is present). We require that the distance $z_{i,i_{p}%
}(t):=x_{i_{p}}(t)-x_{i}(t)$ be constrained so that
\begin{equation}
\setlength{\abovedisplayskip}{2pt}\setlength{\belowdisplayskip}{2pt}
z_{i,i_{p}}(t)\geq\varphi v_{i}(t)+\delta,\text{ \ }\forall t\in\lbrack
t_{i}^{0},t_{i}^{m}], \label{Safety}%
\end{equation}
where $v_{i}(t)$ is the speed of CAV $i\in S(t)$ and $\varphi$ denotes the reaction time (as a rule, $\varphi=1.8$s is used, e.g., \cite{Vogel2003}). If we define $z_{i,i_{p}}$ to be the distance from
the center of CAV $i$ to the center of CAV $i_{p}$, then $\delta$ is a
constant determined by the length of these two CAVs (generally dependent on
$i$ and $i_{p}$ but taken to be a constant for simplicity).

$\textbf{Constraint 2}$ (Safe merging (terminal constraint) between $i$ and
$i-1$): When $i-1 = i_p$, this constraint is redundant since (\ref{Safety}) is enforced, but when $i-1 \ne i_p$ there should be enough safe space at the MP $M$ for a merging CAV to
cut in, i.e.,
\begin{equation}
z_{i,i-1}(t_{i}^{m})\geq\varphi v_{i}(t_{i}^{m
})+\delta. \label{SafeMerging}%
\end{equation}

$\textbf{Constraint 3}$ (Vehicle limitations): There are constraints
on the speed and acceleration for each $i\in S(t)$, i.e.,
\begin{align} 
v_{\min} \leq v_i(t)\leq v_{\max}, \forall t\in[t_i^0,t_i^m],\\
u_{i,\min}\leq u_i(t)\leq u_{i,\max}, \forall t\in[t_i^0,t_i^m],  \label{VehicleConstraints}%
\end{align}
where $v_{\max}>0$ and $v_{\min}\geq0$ denote the maximum and minimum speed
allowed in the CZ, while $u_{i,\min}<0$ and $u_{i,\max}>0$ denote the minimum
and maximum control input, respectively.

$\textbf{Constraint 4}$ (Terminal velocity): In a single CZ problem \cite{xiao2020bridging}, the \emph{optimal} terminal velocity is determined by solving the associated optimal control problem. In contrast, here we impose a constraint $v^m$ determined by the upper (flow control) level:
\begin{equation}
\label{equ:vm}
    v_{i}(t_i^m) = v^m
\end{equation}

\textbf{Problem 1: } Our goal is to determine a control
law {(as well as optimal merging time $t_{i}^{m}$)} to achieve
objectives 1-2 subject to constraints 1-4 for each $i\in S(t)$ governed by the
dynamics (\ref{VehicleDynamics}). The common way to minimize energy
consumption is by minimizing the control input effort $u_{i}^{2}(t)$, noting that the OCBF method
allows for more elaborate fuel consumption models, e.g., as
in \cite{Kamal2013}.
By normalizing travel time and $u_{i}^{2}(t)$, and using $\alpha\in\lbrack0,1)$,
we construct a convex combination as follows: {
\begin{equation}\small
\begin{aligned}J_i(u_i(t),t_i^m; v^m)= \int_{t_i^0}^{t_i^m}\left(\alpha + \frac{(1-\alpha)\frac{1}{2}u_i^2(t)}{\frac{1}{2}\max \{u_{i,\max}^2, u_{i,\min}^2\}}\right)dt \end{aligned}.\label{eqn:energyobja}%
\end{equation}
}If $\alpha=1$, then we solve (\ref{eqn:energyobja}) as a minimum time
problem. Otherwise, by defining $\beta:=\frac{\alpha\max\{u_{i,\max}%
^{2},u_{i,\min}^{2}\}}{2(1-\alpha)}$ and multiplying (\ref{eqn:energyobja}) by
$\frac{\beta}{\alpha}$, we have: {\small
\begin{equation}
\setlength{\abovedisplayskip}{1pt}\setlength{\belowdisplayskip}{1pt}J_{i}%
(u_{i}(t),t_{i}^{m}; v^m):=\beta(t_{i}^{m}-t_{i}^{0})+\int_{t_{i}^{0}}^{t_{i}^{m}%
}\frac{1}{2}u_{i}^{2}(t)dt,\label{eqn:energyobj}%
\end{equation}
}where $\beta\geq0$ is a weight factor that can be adjusted to penalize
travel time relative to the energy cost, {subject to
(\ref{VehicleDynamics}), (\ref{Safety})-(\ref{equ:vm}) and the
terminal constraint $x_{i}(t_{i}^{m})=L$, given $t_{i}^{0},x_{i}(t_{i}%
^{0}),v_{i}(t_{i}^{0})$.}


\subsection{Merging Control: Upper (Flow Control) Level}
The function of the coordinator of CZ $j=1,2,\ldots$ now includes dynamically setting $v^m = v^{m}_j$ in the terminal velocity constraint \eqref{equ:vm}. The traffic in CZ $j$ is influenced by $v^{m}_j$ in two ways: 
$(i)$ CAV $i$'s objective $J_i(u_i(t), t_i^m; v^m_j)$ becomes a function of $v^{m}_j$, and 
$(ii)$ The initial velocity of CAVs in CZ $j$ is exactly the terminal velocity $v^{m}_{j_u}$ in CZ $j$'s upstream CZ, $j_u$. 
Thus, by setting such terminal velocity constraints, the coordinators can control the traffic flow in CZ $j$ based on observations including, but not limited to, the number of CAVs in CZ $j$, $N_j$; the number of CAVs in CZ $j$'s neighboring CZs , $\{N_{j_n}: j_n \in \mathcal{N}(j)\}$; the length of the road segment $L$; and the terminal velocity of the neighboring CZs $v^{m}_{j_n}$.

\textbf{Problem 2} Our goal is to design a terminal velocity constraint controller for CZ $j$
that minimizes the average of CAV objectives: 
\begin{equation}
\label{equ:upper}
\setlength{\abovedisplayskip}{1pt}\setlength{\belowdisplayskip}{1pt}J_j(v^m_{j}(t)) = \frac{1}{N_j(t)} \sum_i^{N_j(t)} J^*_{i}%
(v^m_{j}(t))%
\end{equation}
where $N_j(t)$ denotes the total number of CAVs in CZ $j$ at time $t$ and $J^*_{i}(v^m_{j}(t))$ is the optimal cost in (\ref{eqn:energyobj}). The problem is re-solved at every $t$ when a CAV arrival/departure event occurs that causes a change in $N_j(t)$ and the set of feasible values of $v^m_{j}(t)$ depends on observations from CZ $j$ and its neighboring CZs $j_{n} \in \mathcal{N}(j)$. This is a challenging problem to solve on line. In the sequel, we limit ourselves to a simpler feedback-based policy for $v^m_{j}(t)$, as described in Section \ref{sec:flow_control}.

\section{Hierarchical Control Problem Solution}
\begin{figure}
    \centering
    \includegraphics[width=\linewidth]{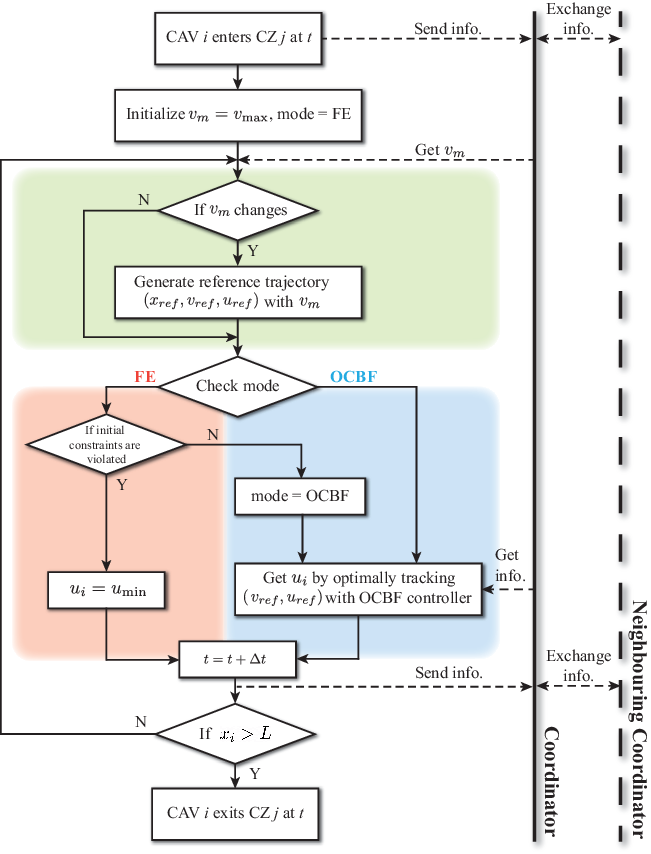}
    \caption{Hierarchical Control Framework over Modular Control Zones. The green, red and blue shaded areas represent the three main components: reference trajectory generation, feasibility enforcement mode, and optimal tracking controller with CBFs.}
    \label{fig:diagram}
\end{figure}

In this section, we provide solutions to \textbf{Problems 1-2} formulated above as shown in Fig. \ref{fig:diagram}. 
\textbf{Problem 1} is solved adopting the OCBF approach \cite{xiao2020bridging}.
\textbf{Problem 2} is solved by using several event-driven methods including a simple fixed $v_m$ and a feedback control method to set the terminal velocity dynamically. 

 
\subsection{\textbf{Problem 1}: Lower Level Merging Control}
The merging control problem can be analytically solved, however, as pointed out in \cite{xiao2020bridging}, it becomes computationally intensive when one of more constraints become active. To solve the merging problem in real time while still \emph{guaranteeing} that all safety constraints are satisfied, we adopt the OCBF approach \cite{xiao2020bridging} through the following steps: 
$(i)$ an optimal control solution for the \emph{unconstrained} optimal control problem (\ref{eqn:energyobj}) is first obtained as a reference control, 
$(ii)$ the resulting reference trajectory is optimally tracked subject to the control bounds \eqref{VehicleConstraints} as well as a set of CBF constraints enforcing \eqref{Safety},\eqref{SafeMerging}. 
Using the forward invariance property of CBFs \cite{xiao2023safe}, these constraints are guaranteed to be satisfied at all times if they are initially satisfied. The importance of CBFs is that they impose \emph{linear} constraints on the control which, if satisfied, guarantee the satisfaction of the associated original constraints that involve the state and/or control. This whole process leads to a sequence of QPs solved over discrete time steps, since the objective function is quadratic and the CBF constraints are linear in the control.

\textbf{Unconstrained optimal control solution}.  
The unconstrained solution refers to the solution of (\ref{eqn:energyobj}) with all safety constraints \eqref{Safety} and \eqref{SafeMerging} inactive. However, the terminal velocity constraint \eqref{equ:vm} is retained as a boundary condition. Using standard Hamiltonian analysis \cite{Bryson1969}, the optimal control, velocity, and position trajectories of CAV $i$ have the form:
\begin{align}
    \label{equ:uc_u}
    u_i(t) &= a_i t + b_i \\
    \label{equ:uc_v}
    v_i(t) &= 1/2 \cdot a_i t^2 + b_i t + c_i \\
    \label{equ:uc_x}
    x_i(t) &= 1/6 \cdot a_i t^3 + 1/2 \cdot b_i t^2 + c_i t + d_i
\end{align}
where the parameters $a_i, b_i, c_i, d_i$ and $t_i^m$ are obtained by solving the following five equations: 
\begin{align}
    \label{equ:ucp_vm_1}
    1/2 \cdot a_i (t_i^0)^2 + b_i t_i^0 + c_i &= v_i^0 \\
    \label{equ:ucp_vm_2}
    1/2 \cdot a_i (t_i^m)^2 + b_i t_i^m + c_i &= v^m \\
    1/6 \cdot a_i (t_i^0)^3 + 1/2 \cdot b_i (t_i^0)^2 + c_i t_i^0 + d_i &= 0 \\
    \label{equ:ucp_vm_4}
    1/6\cdot a_i (t_i^m)^3 + 1/2 \cdot b_i (t_i^m)^2 + c_i t_i^m + d_i &= L \\
    \label{equ:ucp_vm_5}
    \beta - 1/2 \cdot b_i + a_i c_i &= 0
\end{align}
where (\ref{equ:ucp_vm_2}) corresponds to the terminal speed constraint \eqref{equ:vm}.

\textbf{Optimal reference tracking controller with CBFs}: Once we obtain the
unconstrained optimal control solution \eqref{equ:uc_u}-\eqref{equ:uc_x}, we set a reference control 
$u_{ref}(t)=h(u_{i}^{\ast }(t),x_{i}^{\ast }(t),x_{i}(t))$ for some appropriately designed 
$h(u_{i}(\cdot))$ that provides feedback $x_{i}(t)$ from the actual observed CAV trajectory; in the simplest case, we just set 
$u_{ref}(t)=u_{i}^{\ast }(t)$, the solution of the unconstrained problem in (\ref{equ:uc_u}).
We
then design a controller that minimizes $\int_{t_{i}^{0}}^{t_{i}^{m}}\frac{1%
}{2}(u_{i}(t)-u_{\mathrm{ref}}(t))^2dt$ 
subject to all constraints \eqref{Safety} through \eqref{VehicleConstraints}.
This is accomplished by introducing CBFs for the safety constraints to take advantage of their simple form (linear in the control) while still guaranteeing the satisfaction of all constraints through the forward invariance property of CBFs. We review this approach next
(see also \cite{xiao2020bridging}).

First, let $\bm x_{i}(t)\equiv (x_{i}(t),v_{i}(t))$. Based on the vehicle
dynamics \eqref{VehicleDynamics}, define $f(\bm x_{i}(t))=[v_{i}(t),0]^{T}$ and 
$g(\bm x_{i}(t))=[0,1]^{T}$. 
All state constraints, such as \eqref{Safety} and
\eqref{SafeMerging}, can be expressed in the
form $b_{k}(\bm x_{i}(t))\geq 0,k\in \{1,...,B\}$ where $B$ is the number of
constraints. The CBF method maps each constraint $b_{k}(\bm x_{i}(t))\geq 0$
to a new constraint which directly involves the control $u_{i}(t)$ and takes
the (linear in the control) form 
\begin{equation}
L_{f}b_{k}(\bm x_{i}(t))+L_{g}b_{k}(\bm x_{i}(t))u_{i}(t)+\gamma (b_{k}(\bm %
x_{i}(t)))\geq 0,  \label{equ:cbf}
\end{equation}%
where $L_{f},L_{g}$ denote the Lie derivatives of $b_{k}(\bm x_{i}(t))$
along $f$ and $g$ respectively and $\gamma (\cdot )$ denotes some class-$%
\mathcal{K}$ function \cite{xiao2023safe}, usually taken to be linear for simplicity.
The forward invariance
property of CBFs guarantees that a control input that satisfies \eqref{equ:cbf}
will also enforce $b_{k}(\bm x_{i}(t))\geq 0$ at all times. In other words, the safety
constraints \eqref{Safety}, \eqref{SafeMerging} are never violated.

To optimally track the reference speed trajectory, a Control Lyapunov Function (CLF) $V(\bm %
x_{i}(t))$ is used. Letting $V(\bm x_{i}(t))=(v_{i}(t)-v_{\mathrm{ref}}(t))^{2}$, the
CLF constraint takes the form 
\begin{equation}
L_{f}V(\bm x_{i}(t))+L_{g}V(\bm x_{i}(t))u_{i}(t)+\epsilon V(x_{i}(t))\leq
e_{i}(t),  \label{equ:clf}
\end{equation}%
where $\epsilon >0$, and $e_{i}(t)$ is a relaxation variable which makes the
constraint soft. Then, the OCBF controller optimally tracks the reference
trajectory by solving the optimization problem: 
\begin{equation}
\min_{u_{i}(t),e_{i}(t)}\int_{t_{i}^{0}}^{t_{i}^{m}}\left( \beta
e_{i}^{2}(t)+\frac{1}{2}(u_{i}(t)-u_{\mathrm{ref}(t)})^2\right)   \label{equ:ocbf}
\end{equation}%
subject to the CBF constraints %
\eqref{equ:cbf}, the CLF constraints \eqref{equ:clf} and the control bounds \eqref{VehicleConstraints}. 

We can now solve problem \eqref{equ:ocbf} by discretizing $[t_{i}^{0},t_{i}^{m}]$ into
intervals of equal length $\Delta t$ and solving \eqref{equ:ocbf} over each
time interval $[t_{i}^{0} + k\Delta t, t_{i}^{0} + (k+1)\Delta t]$. 
The decision variables $u_{i}(t)$ and $e_{i}(t)$ are assumed
to be constant over each such interval. They can be easily obtained by
solving the following Quadratic Program (QP) problem, since all CBF constraints are
linear in the decision variables $u_{i}(t)$ and $e_{i}(t)$ (fixed over each interval $[t_{i}^{k}, t_{i}^{k} + \Delta t]$): 
\begin{equation}
\label{equ:ocbf_qp}
\begin{split}
\min_{u_{i}(t) ,e_{i}(t)}&~\beta e_{i}(t)^{2}+\frac{1}{2}(u_{i}(t)-u_{\mathrm{ref}}(t))^2 \\
s.t.~&~ \eqref{equ:cbf}, \eqref{equ:clf}, \eqref{VehicleConstraints}, ~~~~~ t = t_i^0 + k\Delta t \\
\end{split}
\end{equation}%

By repeating this process until CAV $i$ exits the CZ, the solution to \eqref{equ:ocbf} is obtained as long as \eqref{equ:ocbf_qp} is
\emph{feasible for each time interval} (this issue is further addressed below). 
Unlike the CBF constraints \eqref{equ:cbf} that guarantee safety, the CLF constraints \eqref{equ:clf} provide convergence to the referenced velocity trajectory, designed so that $v_{\mathrm{ref}}(t_i^m)=v^m$. If convergence to $v^m$ is not fully attained by the time CAV $i$ exits the CZ, this only causes some loss in performance but no violation to any safety constraint.

\textbf{Control and speed reference trajectories.}
The reference trajectories $u_{\mathrm{ref}}(t)$ and $v_{\mathrm{ref}}(t)$ are designed in a position-feedback manner such that $v_{\mathrm{ref}}(t)$ ends with $v^m$, i.e. when $x_i(t) = L$, $v_{\mathrm{ref}}(t_i^m) = v^m$. We introduce a reference time $t_{\mathrm{ref}}$ by solving $x^{\ast}(t_{\mathrm{ref}}) = x(t)$ for any $t$,  
where $x^{\ast}(\cdot)$ is the optimal unconstrained position of a given CAV in \eqref{equ:uc_x}. For computational efficiency, we may use 
an approximate solution $t_{\mathrm{ref}} = t - (x^{\ast}(t) - x(t)) / v^{\ast}(t)$instead. 
Then, we choose the unconstrained optimal trajectory at $t_{\mathrm{ref}}$: 
$v_{\mathrm{ref}}(t) = v^{\ast}(t_{\mathrm{ref}})$, $u_{\mathrm{ref}}(t) = u^{\ast}(t_{\mathrm{ref}})$.

\textbf{Feasibility Enforcement Mode.}
When solving (\ref{eqn:energyobj}) for a single CZ, it is assumed that all safety constraints are initially satisfied.
When OCBF is applied, the additional initial CBF constraints in (\ref{equ:cbf}) are assumed to be satisfied as well. However, when multiple CZs are considered, this assumption is no longer reasonable, as the initial state of a CAV entering a CZ depends directly on the control applied at its upstream CZ.
Thus, a control mechanism is needed to enforce the initial feasibility which can no longer be assumed.

Inspired by the Feasibility Enforcement Zone (FEZ) concept \cite{zhangACC2017}, where initial constraints are enforced in a road segment prior to entering the CZ, we introduce a \emph{Feasibility Enforcement (FE) mode} as shown in Fig. \ref{fig:diagram}. When CAV $i$ enters a CZ at time $t$, it is set to the FE mode with a constant control input of maximum deceleration which enforces the satisfaction of initial constraints as fast as possible up to a maximum distance limit (normally $L/4$) within which feasibility must be attained. 
If none of the constraints is violated, the CAV's mode is switched to \emph{OCBF mode} with the OCBF controller optimally tracking the reference trajectory, and never changed back. 
Of course, the FE mode mechanism does not guarantee the desired initial feasibility, especially when the road length $L$ is short and the initial velocity $v_i^0$ is high. Thus, the coordinator is required to constrain the terminal velocity of CAVs appropriately so as to further enforce feasibility and smooth the traffic flow. In addition, the average time of a CAV in FE mode reflects the robustness of the system: the shorter the average FE mode is in a CZ, the safer and more robust this CZ is.

\textbf{OCBF Control Feasibility.}
As already pointed out, the feasibility of the OCBF controller after each time step in (\ref{equ:ocbf_qp}) is not guaranteed. This is because the time discretization keeps the control constant over a time step and may cause a conflict with the control limits which cannot be predicted by the myopic nature of each QP in the sequence of QPs that are solved.
However, in merging control, we can \emph{enforce this feasibility} by adding a single feasibility guarantee constraint corresponding to each CBF constraint using the method in \cite{xu2022feasibility}.
Specifically, the rear-end CBF constraint (\ref{equ:cbf}) applied to (\ref{Safety}) is $v_{i_p} - v_i - \varphi u_i + k_1b_1(\bm x_i) \geq 0$ and it was shown in \cite{xu2022feasibility} that it can be enforced by any $u_i$ and $k_1>0$ that satisfy
\begin{equation}
u_{i_p} - u_i + k_1(v_{i_p} - v_i - \varphi u_{i, \min}) \geq 0
\label{equ:candidate}
\end{equation}
Similarly, the safe-merging CBF constraint (\ref{equ:cbf}) applied to (\ref{SafeMerging}) is
$v_{i - 1} - v_i - \varphi_2 v_i^2 - \varphi_2 x_iu_i + k_2b_2(\bm x_i) \geq 0$
and it can be shown to be enforced by any $u_i$ and $k_2>0$ that satisfy
\begin{align}
    u_{i - 1} - u_i& - 2\varphi_2 v_iu_i - \varphi_2 v_i u_{\min} \nonumber\\
    &+ k_2(v_{i - 1} - v_i - \varphi_2 v_i^2 - \varphi_2 x_i u_{\min}) \geq 0 \label{equ:candidate2}
\end{align}


\subsection{\textbf{Problem 2}: Upper Level Flow Control}
\label{sec:flow_control}
Flow control is performed by dynamically setting the terminal velocity $v^{m}_j$ for each CZ $j$ in an event-driven way: when a CAV enters or exits CZ $j$ or its neighbors, coordinator $j$ communicates with all its neighboring coordinators $j_n\in \mathcal{N}(j)$ and adjusts $v^{m}_j$ according to accumulated statistical information at the CZ and its neighbors. Coordinator $j$ then broadcasts $v^{m}_j$ to all CAVs within CZ $j$. Once a new $v^m_j$ is received, CAV $i$ generates a new reference trajectory 
through \eqref{equ:uc_u}, \eqref{equ:uc_v}, \eqref{equ:uc_x} and
tracks it with the OCBF controller. This event-driven method avoids frequent rescheduling of CAVs through the lower level control.

The performance of the system is affected by $v^{m}_j$ in two ways. First, $v^{m}_j$ directly influences the CAV objectives at the lower level. Specifically, when the weight $\alpha$ on travel times is relatively large, a large $v^{m}_j$ often results in better performance. On the other hand, an aggressive exit velocity $v^{m}_j$ may result in a higher probability of violating the safety constraints upon entering the downstream CZs, which increases the average time spent in FE mode and consumes more energy as a consequence. 

In this paper, rather than explicitly trying to solve \textbf{Problem 2} through (\ref{equ:upper}), we design a feedback
controller to regulate the terminal velocity $v^{m}_j$ according to the number of CAVs in CZ $j$ and its neighbors which is monitored by each coordinator. The feedback controller has the form:
    \begin{equation}
    \label{equ:feedback_control}
        v^m_j(t) = v_{b} - k N_{j_n}(t)
    \end{equation}
    where $v_{b}$ is a baseline velocity, $k \geq 0$ is a feedback gain parameter, and $N_{j_n}(t)$ is the number of CAVs in the downstream neighboring CZ $j_n$. 
When $k=0$, this becomes a fixed speed controller. A simple, yet effective, selection for $v_{b}$ is the average optimal terminal velocity $v_b = \frac{1}{N_S}\sum_{i \in S}\argmin_{v^m}J_i^*(v^m)$, where where $J^*_{i}(v^m)$ is the optimal cost in (\ref{eqn:energyobj}), $S$ is a set of CAVs sampled with different entry velocity, and $N_S$ is the cardinality of $S$. 

The controller can be modified by replacing $N_{j_n}(t)$ by $N_{j_n}(l,t)$, defined as the number of CAVs in $[0,l]$, the first segment of the CZ with length $l$, since the performance of the downstream CZ is critically affected by the average time in FE mode, therefore, the states of CAVs in this segment is more important than the rest. In this case, $l$ is a parameter to be selected and an alternative selection for $v_b$ is the average velocity of CAVs in the critical zone $[0,l]$, denoted by $\bar v(l)$.
    


\textbf{Efficient trajectory generation when $v^m$ changes.}
Clearly, every time $v^m$ changes in (\ref{equ:ucp_vm_2}), the reference trajectories of all CAVs in the CZ affected by this change need to be re-generated. The event-driven nature of the flow controller above limit the frequency of such changes. Nonetheless, under heavy traffic, this frequency may still be high and motivates an efficient trajectory regenerating method that address this problem.

We first observe that the unconstrained optimal solution to \textbf{Problem 1} \eqref{equ:uc_u}-\eqref{equ:uc_x} is invariant to a time shift, i.e. $u_i(t) = u_i'(t - t')$ where the parameters of $u_i'(t)$ are obtained from \eqref{equ:ucp_vm_1}-\eqref{equ:ucp_vm_5} with $t_i^0 - t'$ replacing $t_i^0$ and $t_i^m - t'$ replacing $t_i^m$. Thus, by shifting time $t'= t_i^0$ forward and replacing $t_m$ by $t_i^m - t_i^0$, \eqref{equ:ucp_vm_1}-\eqref{equ:ucp_vm_5} are equivalent to:
\begin{align}
    c_i &= v_i^0 \label{equ:ucp_shift_vm_1}\\
    1/2 \cdot a_i t_m^2 + b_i t_m + c_i &= v^m \\
    d_i &= 0 \\
    1/6 \cdot a_i t_m^3 + 1/2 \cdot b_i t_m^2 + c_i t_m + d_i &= L \label{equ:ucp_shift_vm_4}\\
    \beta - 1/2 \cdot b_i + a_i c_i &= 0
\end{align}
These can be combined into a single cubic equation in $t_m$ (for notational simplicity, $v_i^0$ and $v^m$ are replaced with $v_0$ and $v_m$ respectively in the sequel):
{\small
\begin{equation}
    \label{equ:tm}
    \beta t_m^3 + (2 v_0 + v_m)t_m^2 + (-3L + 6v_0v_m + 6v_0^2)t_m - 12L v_0 = 0
\end{equation}
}
Denoting the left side of \eqref{equ:tm} by $f(t_m, v_m)$, we get 
{\small
\begin{align}
\frac{\partial f(t_m, v_m)}{\partial v_m} &= t_m^2 + 6v_0t_m \nonumber\\
\frac{\partial f(t_m, v_m)}{\partial t_m} &= 3\beta t_m^2 + (4v_0 + 2v_m)t_m - 3L + 6v_0v_m + 6v_m^2 \nonumber
\end{align}
}
Considering a small perturbation in $v_m$, we have $f(t_m, v_m + \Delta v) = f(t_m, v_m) + \frac{\partial f(t_m, v_m)}{\partial v_m}\Delta v$, which results in a perturbation in $t_m$ such that
$f(t_m + \Delta t, v_m + \Delta v) = f(t_m, v_m) = 0$.
Then, using a first-order Taylor expansion,
{\small
\begin{equation}
\label{equ:perturb_tm}
    f(t_m + \Delta t, v_m + \Delta v) = f(t_m, v_m + \Delta v) + \frac{\partial f(t_m, v_m)}{\partial t_m} \Delta t
\end{equation}
}
which yields
{\small
\begin{equation}
\label{equ:delta_t}
    \begin{split}
        \Delta t &= -\left(\frac{\partial f(t_m, v_m)}{\partial v_m}\right) \left/ \left(\frac{\partial f(t_m, v_m + \Delta v)}{\partial t_m}\right)\Delta v\right. \\
        &= -\frac{t_m^2 + 6v_0t_m}{3\beta t_m^2 + (4 v_0 + 2 v_m')t_m - 3L + 6v_0v_m' + 6v_0^2}\Delta v
    \end{split}
\end{equation}
}
where $v_m' = v_m + \Delta v$ is the perturbed $v_m$. This provides an analytical expression of the perturbed terminal time $t_m' = t_m + \Delta t$ when $v_m$ is perturbed by $\Delta v$. 

Now the problem turns into finding the unconstrained optimal solution to \textbf{Problem 1} with fixed terminal time $t_m'$. The solution shares the form \eqref{equ:uc_u}-\eqref{equ:uc_x} with parameters obtained 
by solving \eqref{equ:ucp_vm_1}-\eqref{equ:ucp_vm_4}. By similarly shifting time $t'= t_i^0$ forward and replacing $t_m$ by $t_i^m - t_i^0$, \eqref{equ:ucp_vm_1}-\eqref{equ:ucp_vm_4} are equivalent to \eqref{equ:ucp_shift_vm_1}-\eqref{equ:ucp_shift_vm_4}
which has a unique analytical solution if $t_m \neq 0$ (which obviously holds):
\begin{equation}
\label{equ:ucp_vm_tm}
    \begin{split}
        a_i &= 6/t_m^2(v^m - v_i^0) - 12/t_m^3(L - v_i^0t_m) \\
        b_i &= -2/t_m(v^m - v_i^0) + 6/t_m^2(L - v_i^0t_m) \\
        c_i &= v_i^0, ~~~~~d_i = 0
    \end{split}
\end{equation}
Thus, every time $v^m$ is changed by $\Delta v$, we can efficiently generate the unconstrained optimal trajectories by setting $t_m' = t_m + \Delta t$ using \eqref{equ:delta_t} and then using $t_m'$ into \eqref{equ:ucp_vm_tm} to update all the parameters without re-solving \eqref{equ:ucp_vm_1}-\eqref{equ:ucp_vm_5}.

\section{Simulation Results}
In this section, we use the multi-model traffic flow simulation platform Vissim to construct the two-CZ configuration in Fig. \ref{fig:module_CZ}. We replicate this model in a Python-based simulation with the same traffic input and apply the lower level controller only (without flow control using terminal velocity constraints) as a \emph{baseline} to evaluate the traffic performance in two consecutive merging roadways and compare it to the performance obtained using our proposed modular control zone framework. In the simulations, only the terminal velocity $v^{m}_1$ of CZ1 is set for flow control, while $v^{m}_2$ is set to be the average exiting velocity of CAVs passing CZ2 with the lower level controller only for fair comparison.

\begin{figure}[htb]
\vspace{-10pt}
    \centering
    \includegraphics[width=\linewidth]{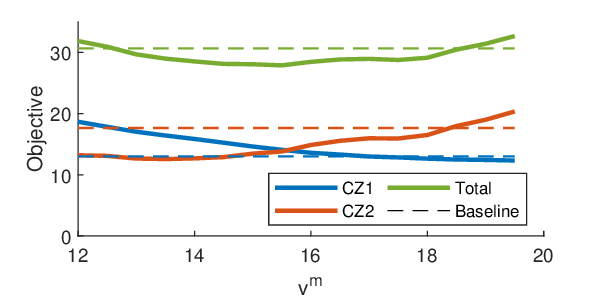}
    \vspace*{-16pt}
    \caption{Objectives under flow control with different fixed $v^{m}_1$}
    \label{fig:diff_v}
\end{figure}

\textbf{Flow control with fixed $v^m$ vs. No flow control.}
We start with the flow controller (\ref{equ:feedback_control}) with fixed $v^m$. The parameter settings are as follows: $L_1 = L_2 = 200m, \delta = 0m, \varphi = 1.8s$, $v_{\max }=30m/s$, $v_{\min }=0$, $u_{\max } = -u_{\min } =4m/s^{2}$, $\alpha = 0.0625$, $v^{m}_2 = 18.5m/s$. We use a fixed traffic rate $400$ CAVs/h in each road segment and simulate around 120 CAVs. Simulations are performed using a fixed $v^{m}_1$ value, which varies over $12-20m/s$ with step size $0.5m/s$. The objective of each CZ as well as the total objective corresponding to different $v^{m}_1$ are shown in Fig. \ref{fig:diff_v} where dashed lines show the baseline.
The figure shows that a higher $v^{m}_1$ results in a better performance in CZ1, but worse in CZ2, which well aligns with our expectation as explained in Sec. \ref{sec:flow_control}. The speed $v^{m}_1$ constrains the aggressive exit velocity which has a high probability of violating the safety constraints when entering CZ2; thus, it improves the performance in CZ2 when $v^{m}_1$ is low. The total objective can be approximated by a quadratic function with the minimum value occurring when $v^{m}_1$ is around $15m/s$.

The effectiveness of the flow controller with fixed $v^{m}_1$ can also be illustrated in terms of the average time CAVs spend in FE mode (FEm) entering CZ2. We show the key feasibility metrics in Table \ref{tab:fem_time} including the total number of CAVs with FE mode active, the average active FE mode time, and the total number of CAVs with infeasible QPs. 
Both FEm count and average FEm time drop when flow control is exerted with fixed $v^{m}_1$ by 25.8\% and 60.5\% respectively. Since we have already included the feasibility guarantee constraints \eqref{equ:candidate} and \eqref{equ:candidate2} in the OCBF controller, the only reason for the existence of infeasible QPs using no flow control is that the \emph{initial} feasibility fails to be enforced within the FE mode distance limitation. Both the initial feasibility and the QP feasibility of CAV are more easily enforced with fixed $v^{m}_1$ flow control, thus making the system safer and more robust.
\begin{table}[htb]
    \caption{Feasibility Comparison of Flow Control vs. no flow control}
    \centering
    \begin{tabular}{cccc}
        \toprule[1pt]
         & FEm count & FEm Time (s) & Infeasible Count\\
        \midrule
        No flow control & 62 & 0.86 & 1\\
        $v^{m}_1=15m/s$ & 46 & 0.34 & 0\\
        $v^{m}_1=12m/s$ & 26 & 0.15 & 0\\
        \bottomrule[1pt]
    \end{tabular}
    \label{tab:fem_time}
\end{table}

\textbf{Feedback control on $v^m$.}
We apply a simple feedback controller as in \eqref{equ:feedback_control}
with the same parameter settings as before under a fixed traffic rate of 400 CAVs/h and parameters $v_{b} = 18m/s$, $k = 1/2$ and $l = L/4$. We compare the performance of the feedback controller with the fixed value flow controller ($v^{m}_1 = 15m/s$) and the baseline using only the lower level controller as shown in Table. \ref{tab:feedback1}.
\begin{table}[htb]
    \centering
    \caption{Performance Comparison}
    \label{tab:feedback1}    
    \begin{tabular}{ccccc}
    \toprule
         & Metric & Feedback Control & Fixed $v^m$ & No flow control \\
    \midrule
        \multirow{3}{*}{CZ1} & t & 12.45 & 13.44 & 10.65\\
         & e & 4.51 & 5.66 & 5.81 \\
         & obj. & 12.81 & 14.62 & 12.92 \\
    \midrule
        \multirow{3}{*}{CZ2} & t & 10.31 & 12.11 & 10.23\\
         & e & 8.32 & 5.38 & 13.31 \\
         & obj. & 15.19 & 13.46 & 20.13 \\
    \midrule
        Total & obj. & 28.00 & 28.07 & 33.03 \\
    \bottomrule
    \end{tabular}
\end{table}

Table \ref{tab:feedback1} shows that the flow controller with fixed $v^m$ and feedback control on $v^m$ both outperform the baseline in the sense of total objective, with approximately 15\% improvement. CAVs using only the lower level controller always have a shorter travel time than the modular CZ method with flow control at the expense of an energy consumption rise in CZ2 causing a total performance loss. There is no significant performance increase using the feedback controller due to the constant traffic rate used in the simulation. As can be seen in Fig. \ref{fig:diff_v}, $v^{m}_1 = 15m/s$ is already close to optimal, leaving little room for extra improvement. However, a feedback controller pays off when the traffic rate increases over time.

\begin{figure}[htb]
    \centering
    \includegraphics[width = \linewidth]{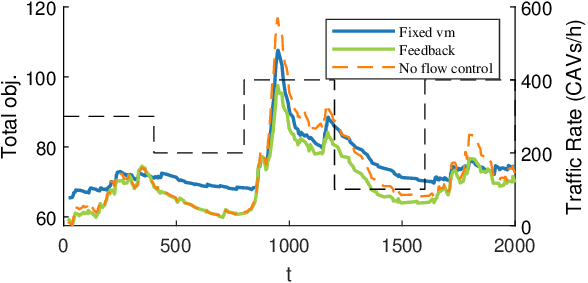}
    \vspace*{-16pt}
    \caption{Performance comparison of flow control with fixed $v^m$ vs. feedback control on $v^m$ vs. no flow control under changing traffic rate. The changing traffic rate is shown in the dashed black line.}
    \label{fig:rate}
\end{figure}

\textbf{Varying traffic input.}
We construct a varying traffic input with rates $\{300, 200, 400, 100, 400\}$ CAVs/h in time interval $\{[0, 400], [400, 800], [800, 1200], [1200, 1600], [1600, 2000]\}$ respectively. The same parameter settings are used except $\alpha = 0.25$ to put more emphasis on travel time. The flow controller with fixed value $v^{m}_1=15m/s$ and the feedback controller $v^{m}_1 = \bar v + 2 - 1/2 n(L/4)$ are both applied and compared with the baseline of no flow control. The total objective of each CAV is shown in Fig. \ref{fig:rate} with the exponential smoothing technique used for visualization. CAVs using flow control with fixed $v^m$ (solid blue line) outperform those using only lower-level control (dashed orange line) when the traffic rate is high. However, the fixed $v^m$ constraint becomes too conservative under low traffic, thus restricting CAV performance. However, by applying feedback flow control (solid green line), CAVs can achieve a similar near-optimal performance as using only the lower level control under low traffic rate and outperform it when the system becomes congested.  

The numerical results in Table \ref{tab:feedback} also illustrate the benefit of feedback control on $v^m$: it constrains the aggressive exit velocity of CAVs in heavy traffic and adjusts $v^m$ according to traffic conditions to avoid being too conservative, thus outperforming both no flow control and fixed $v^m$. The performance increase is not as significant as in Table \ref{tab:feedback1} mainly due to $\alpha=0.25$ giving more emphasis on travel time, thus favoring the aggressive no flow control method.
\begin{table}[htb]
    \centering
    \caption{Performance Comparison}
    \label{tab:feedback}    
    \begin{tabular}{ccccc}
    \toprule
         & Metric & Feedback Control & Fixed $v^m$ & No flow control \\
    \midrule
        \multirow{3}{*}{CZ1} & t & 10.69 & 12.48 & 10.66\\
         & e & 8.45 & 5.58 & 5.81 \\
         & obj. & 36.96 & 38.85 & 34.23 \\
    \midrule
        \multirow{3}{*}{CZ2} & t & 9.72 & 10.69 & 10.23\\
         & e & 9.88 & 9.41 & 13.31 \\
         & obj. & 35.82 & 37.92 & 40.59 \\
    \midrule
        Total & obj. & 72.78 & 76.77 & 74.72 \\
    \bottomrule
    \end{tabular}
\end{table}

\section{Conclusion and Future Work}
We have presented a modular control zone framework consisting of decentralized optimal control of CAVs in the lower level and flow control in the upper level with he goal of scaling up optimal and safe controllers from a single CZ to an entire network.
CZs are interconnected through the terminal velocity parameter $v^m$ as the interface that regulates the CZ exit speed of CAVs using a decentralized OCBF controller. A fixed $v^m$ and feedback control method are proposed to set $v^m$ for flow control at the upper level. 
Simulations of two consecutive CZs for merging roadways show a significant improvement compared to a baseline of no flow control, especially under changing traffic conditions. 
Future work is targeting the use of this framework in mixed traffic (CAVs co-existing with human-driven vehicles) and improved upper level flow control methods.

\bibliographystyle{ieeetr}
\bibliography{cav}
{}

\end{document}